\documentclass[aps,prd,preprint,tightenlines,superscriptaddress,showpacs,byrevtex]{revtex4}
\usepackage{graphicx}
\usepackage{dcolumn}
\usepackage{bm}

\graphicspath{{./}}
\def\bz {B^{0}}
\def\bp {B^{+}}

\def\dz {D^{0}}
\def\dbar {\overline{D}{}^{0}}
\def\jpsi {J/\psi}
\def\pip  {\pi^{+}}
\def\pim {\pi^{-}}

\def\Mbc {M_{\rm bc}}    

\def\row {\rightarrow}
\def\Dmb {\Delta M_{B}}
\def\Md  {M_{K\pi}}      

\def\bzdecay {\bz \to \jpsi \dbar}
\def\bpdecay {\bp \to \jpsi \dbar \pip}

\begin{document}
\vspace*{-3\baselineskip}
 \resizebox{!}{3cm}{\includegraphics{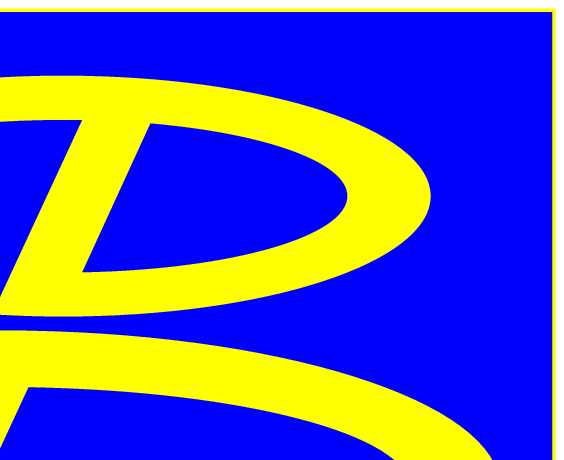}}
\preprint{\vbox{
                 \hbox{KEK Preprint 2005-1}
                 \hbox{Belle Preprint 2005-11}
}}

\title{\boldmath Search for $\bz \row\jpsi \dbar$ and $\bp \row \jpsi \dbar \pip$ decays }
\affiliation{Budker Institute of Nuclear Physics, Novosibirsk}
\affiliation{Chiba University, Chiba}
\affiliation{Chonnam National University, Kwangju}
\affiliation{University of Cincinnati, Cincinnati, Ohio 45221}
\affiliation{University of Hawaii, Honolulu, Hawaii 96822}
\affiliation{High Energy Accelerator Research Organization (KEK), Tsukuba}
\affiliation{Hiroshima Institute of Technology, Hiroshima}
\affiliation{Institute of High Energy Physics, Chinese Academy of Sciences, Beijing}
\affiliation{Institute of High Energy Physics, Vienna}
\affiliation{Institute for Theoretical and Experimental Physics, Moscow}
\affiliation{J. Stefan Institute, Ljubljana}
\affiliation{Kanagawa University, Yokohama}
\affiliation{Koreab University, Seoul}
\affiliation{Kyungpook National University, Taegu}
\affiliation{Swiss Federal Institute of Technology of Lausanne, EPFL, Lausanne}
\affiliation{University of Ljubljana, Ljubljana}
\affiliation{University of Maribor, Maribor}
\affiliation{University of Melbourne, Victoria}
\affiliation{Nagoya University, Nagoya}
\affiliation{Nara Women's University, Nara}
\affiliation{National Central University, Chung-li}
\affiliation{National United University, Miao Li}
\affiliation{Department of Physics, National Taiwan University, Taipei}
\affiliation{H. Niewodniczanski Institute of Nuclear Physics, Krakow}
\affiliation{Nihon Dental College, Niigata}
\affiliation{Niigata University, Niigata}
\affiliation{Osaka City University, Osaka}
\affiliation{Osaka University, Osaka}
\affiliation{Panjab University, Chandigarh}
\affiliation{Peking University, Beijing}
\affiliation{Princeton University, Princeton, New Jersey 08544}
\affiliation{Saga University, Saga}
\affiliation{University of Science and Technology of China, Hefei}
\affiliation{Seoul National University, Seoul}
\affiliation{Sungkyunkwan University, Suwon}
\affiliation{University of Sydney, Sydney NSW}
\affiliation{Tata Institute of Fundamental Research, Bombay}
\affiliation{Toho University, Funabashi}
\affiliation{Tohoku Gakuin University, Tagajo}
\affiliation{Tohoku University, Sendai}
\affiliation{Department of Physics, University of Tokyo, Tokyo}
\affiliation{Tokyo Institute of Technology, Tokyo}
\affiliation{Tokyo Metropolitan University, Tokyo}
\affiliation{Tokyo University of Agriculture and Technology, Tokyo}
\affiliation{University of Tsukuba, Tsukuba}
\affiliation{Virginia Polytechnic Institute and State University, Blacksburg, Virginia 24061}
\affiliation{Yonsei University, Seoul}
 \author{L.~M.~Zhang}\affiliation{University of Science and Technology of China, Hefei} 
   \author{Z.~P.~Zhang}\affiliation{University of Science and Technology of China, Hefei} 
   \author{K.~Abe}\affiliation{High Energy Accelerator Research Organization (KEK), Tsukuba} 
   \author{K.~Abe}\affiliation{Tohoku Gakuin University, Tagajo} 
   \author{I.~Adachi}\affiliation{High Energy Accelerator Research Organization (KEK), Tsukuba} 
   \author{H.~Aihara}\affiliation{Department of Physics, University of Tokyo, Tokyo} 
   \author{Y.~Asano}\affiliation{University of Tsukuba, Tsukuba} 
   \author{T.~Aushev}\affiliation{Institute for Theoretical and Experimental Physics, Moscow} 
   \author{S.~Bahinipati}\affiliation{University of Cincinnati, Cincinnati, Ohio 45221} 
   \author{A.~M.~Bakich}\affiliation{University of Sydney, Sydney NSW} 
% ...
\collaboration{The Belle Collaboration}


\begin{abstract}
We report
the results of a search for the decay
modes $\bz \row\jpsi \dbar$ and $\bp \row \jpsi \dbar \pip$. The analysis is
based on 140 $\rm fb^{-1}$ of data accumulated by the Belle detector at the
KEKB asymmetric-energy $e^{+} e^{-}$ collider.
No significant signals are observed and we determine
the branching fraction upper limits ${\cal B} (\bz \row \jpsi \dbar) < 2.0
\times 10^{-5}$ and ${\cal B} (\bp \row \jpsi \dbar \pip) < 2.5 \times
10^{-5}$ at 90\% confidence level.
These results rule out the explanation of the excess in the low momentum region of the inclusive
$\jpsi$ spectrum as intrinsic charm content in the $B$ meson. The branching fractions of the
corresponding nonresonant decay
 channels are also reported.

\end{abstract}

\pacs{13.25.Hw,14.40.Nd} \maketitle

The inclusive spectrum of $B \row \jpsi + X$ has been 
studied extensively and
is consistent with the prediction of nonrelativistic QCD
calculations \cite{NRQCD}, except for an excess in the low momentum
region \cite{1,2}.
This momentum region
corresponds to the $\jpsi$ meson recoiling against
particle systems with an invariant mass of
$\sim 2$ GeV/$c^2$. The observed excess below $0.8$ GeV/$c$ corresponds to a branching fraction of
a few times $10^{-4}$.  

Several hypotheses \cite{BN,3,5}  have been proposed to explain this excess.
One of the decay modes proposed in Ref. \cite{BN},
$\bp \row \jpsi \Lambda \overline{p}$ \cite{conjugate},
has been studied by BaBar \cite{4} and Belle \cite{zsl}.
The measured branching fraction, of order $10^{-5}$,
is too small to account for the excess.  

Chang and Hou \cite{3} proposed intrinsic charm ($c\bar{c}$) in the $B$ meson as an explanation. The
intrinsic charm pair transforms into a $c\bar{c}$ final state when the $B$ meson decays. The most
promising decay modes are $\bz(d\bar{b}c\bar{c})\row \jpsi \overline{D}{}^{(*)0}$ and
$\bp(u\bar{b}c\bar{c})\row \jpsi \dbar \pip$.
According to this hypothesis, if the intrinsic charm content of the $B$ is not much less than $1\%$,
the branching fractions of the above decay modes could be $\sim 10^{-4}$, while the prediction from the
standard QCD framework is of order $10^{-8}$ \cite{5}.

In this paper, we report on a search for the decay modes $\bz \row\jpsi \dbar$ and $\bp \row
\jpsi \dbar \pip$. The analysis is based on a data sample of 140 $\rm fb^{-1}$,
which contains 152 $\times 10^6$ $B \overline B$ pairs,
accumulated at the $\Upsilon$(4$S$) resonance with the Belle detector
\cite{belle} at the KEKB 8 GeV $e^-$ and 3.5 GeV $e^+$ asymmetric collider
\cite{kekb}.

The Belle detector is a large-solid-angle magnetic spectrometer that consists
of a three-layer silicon vertex detector (SVD), a 50-layer central drift
chamber (CDC), an array of aerogel threshold \v{C}erenkov counters (ACC), a
barrel-like arrangement of time-of-flight scintillation counters (TOF), and an
electromagnetic calorimeter comprised of CsI(Tl) crystals (ECL). These
detectors are located inside a superconducting solenoid coil that provides a
1.5 T magnetic field. An iron flux-return located outside of the coil is
instrumented to detect $K_L$ mesons and to identify muons (KLM).

Events are required to pass the hadronic event selection criteria
\cite{PRD03}.
To suppress continuum backgrounds ($e^+ e^- \to q\bar{q}$, where $q = u,\,d,\,s,\,c$), we require
$R_{2}<0.5$, where $R_{2}$ is the ratio of the second to zeroth Fox-Wolfram moments \cite{FW}.

The selection criteria for $\jpsi$ mesons decaying to $l^+l^-$ (where
$l=e,\,\mu$) are identical to those used in Ref. \cite{zsl,PRD03}. To remove
charged particle tracks that do not come from the interaction region,
we require that the leptons originate from within 5 cm of the nominal
interaction point (IP) along the beam direction.
Both tracks are required to be positively identified as leptons. In order to reduce the effect of
bremsstrahlung or final state radiation, clusters detected in the ECL within $0.05$ radians of the
original $e^-$ or $e^+$ direction are added in the invariant mass calculation.
The $\jpsi$ candidate is required to satisfy an
asymmetric invariant mass requirement that takes account of
the radiative tail: $-150(-60)<M_{e^+e^-(\gamma)}(M_{\mu^+\mu^-})-m_{\jpsi}
<36(36)$ MeV/$c^2$,
where $m_{\jpsi}$ is the nominal $\jpsi$ mass \cite{pdg}.
In order to
improve the momentum resolution, vertex and mass constrained fits are
then applied to the $\jpsi$ candidates that pass  
the above selection criteria.

For charged pion and kaon identification, the specific ionization ($dE/dx$) in the CDC, the flight time
measured in the TOF, and the response of the ACC are combined into a likelihood $L_{h}$, where $h$
stands for the hadron type ($\pi$, $K$, $p$). A track is labeled as a kaon if
$L_{K}/(L_{K}+L_{\pi})>0.5$ or a pion if $L_{\pi}/(L_{\pi}+L_{K})>0.3$; the respective efficiencies are
90\% and 92\%, while the respective $\pi$/$K$ misidentification rates are 10\% and 13\%. Tracks in the
kaon sample with $L_{p}/(L_{p}+L_{K})>0.99$ are reclassified as protons and thereby discarded. All
tracks positively identified as electrons are rejected.

A $\dz$ meson candidate is reconstructed from a $K^-$ and a $\pi^+$ meson pair. The two
tracks must satisfy $dr<0.3$ cm 
and $|dz|<5$ cm, where $dr$ ($dz$) is the impact parameter perpendicular to (along) the beam direction
with respect to the IP, determined run-by-run.
We select $\dz$ candidates for further analysis within the mass window
$|\Md-m_{\dz}|<150$ MeV/$c^2$, where
$m_{\dz}$ is the nominal $\dz$ mass \cite{pdg}. We also apply a vertex
constrained fit to the $\dz$ candidates.

For the charged pion in $\bpdecay$, we apply the requirements with looser $dr < 0.6$ cm, $|dz| < 5$ cm,
and tighter $L_{\pi}/(L_{\pi}+L_{K})>0.9$ due to its low momentum.

$B$ mesons are reconstructed by combining a $\jpsi$ and a $\dbar$ candidate
for $\bzdecay$, and an additional pion with the same charge as the kaon
from $\dbar$ for $\bpdecay$.
To reduce combinatorial background, we impose a requirement on 
the quality ($\chi^2$) of the vertex fit for the leptons from $\jpsi$ and the $\dbar$ trajectory (and
the $\pip$ for the $\bp$ case).
The vertexing requirement retains 94\% (78\%) of the $\bz$ ($\bp$) signal.

We reject $\bp \to \psi(2S) K^+$ [$\psi(2S) \to \jpsi \pi^+\pi^-$] decay
by requiring $M_{l^+l^-(\gamma)\pi^+\pi^-}-M_{l^+l^-(\gamma)}$ to be
outside of the $\pm 15$ MeV/$c^2$ window around the nominal mass difference
between $\psi(2S)$ and $\jpsi$.
We require $|\cos{\theta_{B}|}<0.8$ to further suppress combinatorial
background,
where $\theta_B$ is
the angle between the $B$
flight direction and positron beam direction in the center-of-mass
(cms) frame.


We select $B$ candidates
by requiring that the beam-energy constrained mass ($\Mbc\equiv\sqrt{E_{\rm
beam}^{2}-P^{2}_{B}}$) and the  the mass difference ($\Dmb\equiv M_{B} - m_B$) \cite{dmb} lie within
the region  $\Mbc>5.2$ GeV/$c^{2}$ and $-0.3 < \Dmb <0.2$ GeV/$c^2$; here, $E_{\rm beam}$ and $P_{B}$
are the beam energy and $B$ momentum in the cms, while $M_{B}$ and $m_B$ are the reconstructed and the
nominal mass of the $B$ meson.
The signal region 
for $\bz$ ($\bp$) candidates is defined as
$5.27<\Mbc<5.29$ GeV/$c^{2}$, $|\Dmb|<$ 19.0 (17.1) MeV/$c^{2}$, $|\Md-m_{\dz}|<16.5$ MeV/$c^{2}$,
corresponding to three standard deviation windows based on Monte Carlo (MC) simulation
and the data control samples described later. The candidates outside of the signal region are used to
determine the background components in the fit described below.

After the selection, around 3.6\% (42\%) of $\bz$ ($\bp$) candidate events have more
than one $B$ candidate.
In multiple candidate cases,
we select the candidate with the best vertex fit quality.

Background is divided into categories that have distinct shapes in the ($\Mbc, \Dmb, \Md$)
distributions: nonresonant background, two types of combinatorial backgrounds, and peaking backgrounds.
In the case of nonresonant background, a $B$ meson decays to the signal final state, but the $K^+$ and
$\pi^-$ mesons do not come from a $\dbar$ decay.
In the case of combinatorial background, the reconstructed $\jpsi$ and $\dbar$ [and a pion, for a $\bp$
candidate] come from different $B$ mesons (90\% of the time) or from continuum events (10\%).
No peak appears in the ($\Mbc, \Dmb$) distribution; however, one subclass---cmb(D0)---has a peak in the
$\Md$ distribution ($\dbar$ correctly reconstructed), while the other---cmb(fake D0)---does not (fake
$\dbar$).

The peaking background shows an enhancement in the $\Mbc$ signal region.
For the $\bz \row \jpsi \dbar$ signal, one source of this background is from $\bp \row \psi(2S) (\chi_{c})
K^{+}$ decay, where the $\jpsi$ and $K^{+}$ are combined
with a pion from the $B^-$ meson. Another source is from
$\bp \to \jpsi K^{+} \pim\pip$ ($\bz \to \jpsi K^{+} \pim \pi^{0}$) decay where the second pion is missed.
For the $\bp \row \jpsi \dbar\pip$ signal,
one source of peaking background comes from the aforementioned $\bp$($\bz$) decay with the $\pi^\pm$
($\pi^0$) replaced by a charged pion from the other $B$ meson. Another is from the $\bz \row \jpsi
K^{+} \pim$ decay combined with a $\pip$ from the accompanying $B$. For the $\bz$ ($\bp$) signal, the
first-mentioned peaking background distributes broadly around the $\Dmb$ signal region, while the
second exhibits a narrow peak shifted to negative (positive) values by a pion mass.
The narrow
peak is excluded by limiting the fit
region to $\Dmb > -0.12$ ($<+0.12$) GeV/$c^2$.

The yields are extracted by maximizing the three-dimensional (3D)
extended likelihood function,
\[
{\cal L} = \frac{e^{-\sum\limits_{k}N_{k}}}{N!}\prod^N_{i=1}\left[\sum_{k}N_{k}\times P_k(M_{{\rm
bc}}^{i},\Delta M_B^{i},M_{K\pi}^{i})\right],
\]
where $N$ is the total number of candidate events,
$i$ is the identifier of the $i$-th event,  
$N_{k}$ and $P_{k}$ are the
yield and  probability density function (PDF) of component $k$,
which corresponds to  
the signal and
each aforementioned background.

The signal PDF is determined using MC simulation and control-sample data.
A Gaussian is
used as the $\Mbc$ PDF. Since $\Dmb$ is correlated with $\Md$ but
not with $\Dmb-\Md$,  
we use the product of
the $\Md$ and $\Dmb-\Md$ PDFs, each of which is the sum of two Gaussians.
The parameters of the $\Md$ PDF are extracted
from the inclusive $\dz$ data
sample with cms momentum less than 1.5 GeV/$c$, using the same $\dz$ selection
criteria as those for $B$ decays.   
The mean and width of the $\Mbc$ PDF and the main Gaussian of the $\Dmb$ PDF
(integrated over $\Md$) are calibrated using a
$\bz \row \jpsi K^{*0} (K^{*0} \to K^+\pim)$ control sample.

The $\bp \row \jpsi \dbar \pip$ signal PDF has an additional component where
a low momentum pion is incorrectly assigned as the pion from $B$ decay.
We construct its PDF from the product of the double-Gaussian in
$\Md$ with a two-dimensional (2D) smoothed histogram in
$\Mbc$ and $\Dmb$.
The fraction of this component   
is estimated to be ($37.4\pm0.7$)\% from signal MC and is fixed in the fit.

For nonresonant background, the $\Mbc$ and $\Dmb$ PDFs are taken to be the
same as the signal PDFs integrated over $\Md$,
while a second-order polynomial is used for the $\Md$ PDF.
For combinatorial background, 
a threshold function \cite{ARGUS}
 is used for
the $\Mbc$ PDF.
For $\bzdecay$, a first-order polynomial is used for the $\Dmb$ PDF.
To take into account the kinematic boundary for $\bpdecay$, we use
another threshold function,
\[
     P_{\rm thr}(x,x_c; p,c) = \left\{
   \begin{array}{ll}
     (x - x_c)^p e^{-c(x-x_c)} & (x \ge x_c) \\
     0                         & (x < x_c) \\
   \end{array}\right.
\]
with $x = \Dmb$ and $ x_c=\Md - (m_{\bp}-m_{\jpsi}-m_{\pip})$,
where $m_{\bp}$ and $m_{\pip}$ are the nominal $\bp$ and $\pip$ masses.
%
The same $\Md$ PDF as used for signal is used for cmb(D0), while a first-order polynomial is used for
cmb(fake D0).
For peaking backgrounds, the PDFs are modeled by 3D smoothed   
histograms from a large $\jpsi$ inclusive MC sample.
%

In the fit, the value of $N_k$ and the parameters for the polynomials and threshold functions are
allowed to float.


\begin{figure}[!hbtp]
\begin{center}
\includegraphics[width=0.4\textwidth]{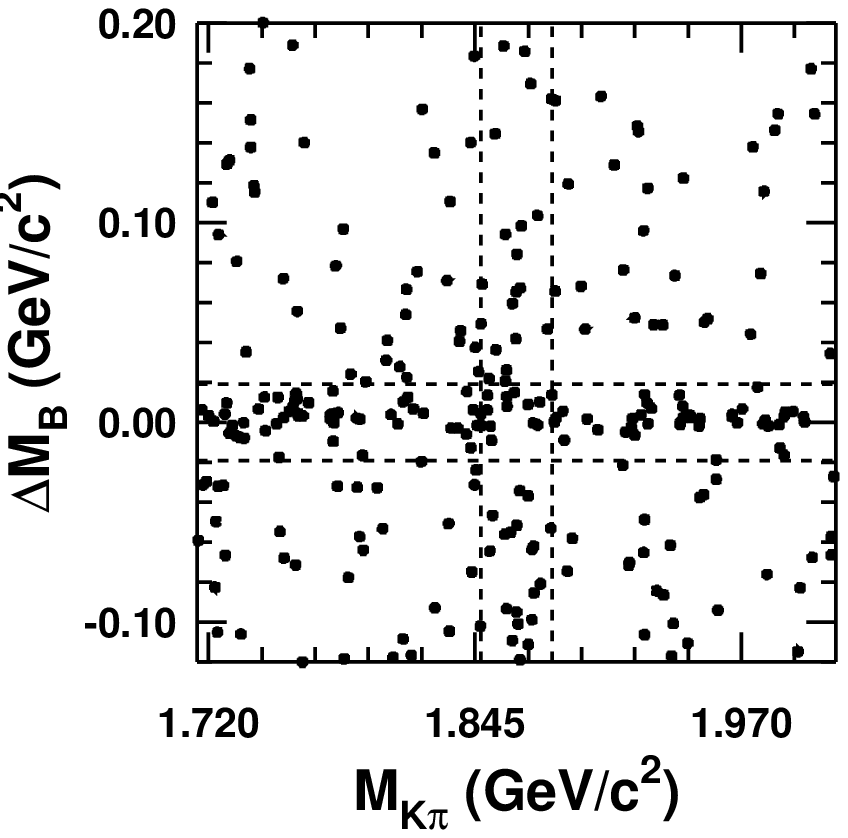}%
\includegraphics[width=0.4\textwidth]{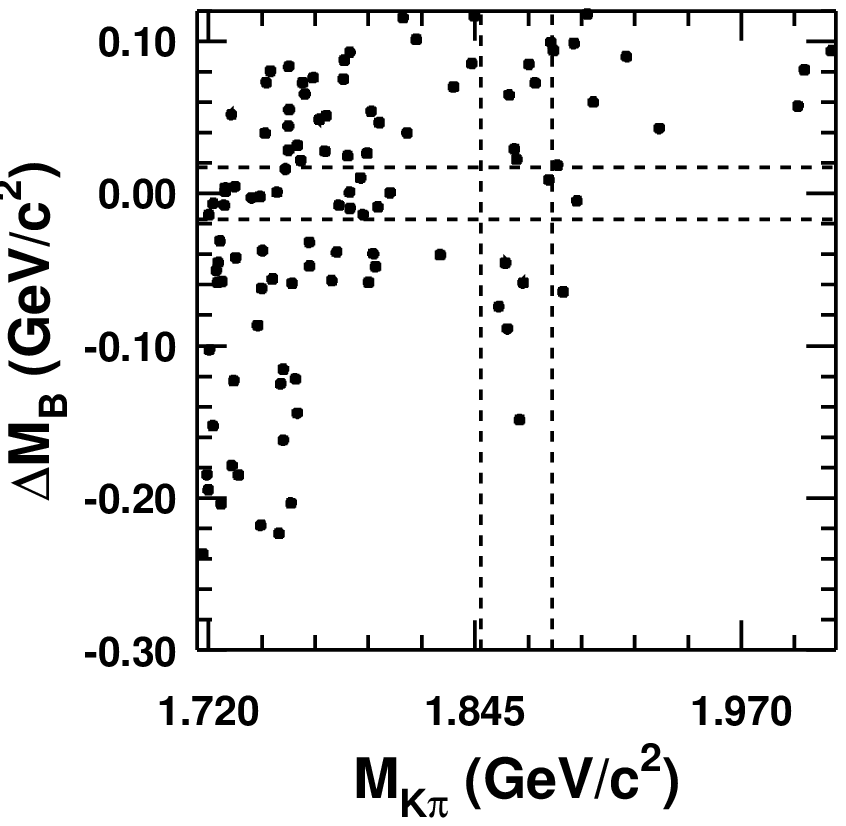}
\caption{\label{sbox}
 ($\Md$, $\Dmb$) scatter plots for data in the $\Mbc$
 signal region for $\bzdecay$ (left) and $\bpdecay$ (right). Dashed lines indicate
 the signal regions for $\Md$ and $\Dmb$. }
\end{center}
\end{figure}

\begin{figure*}[!hbtp]
\begin{center}
\includegraphics[width=0.33\textwidth,height=0.35\textheight]{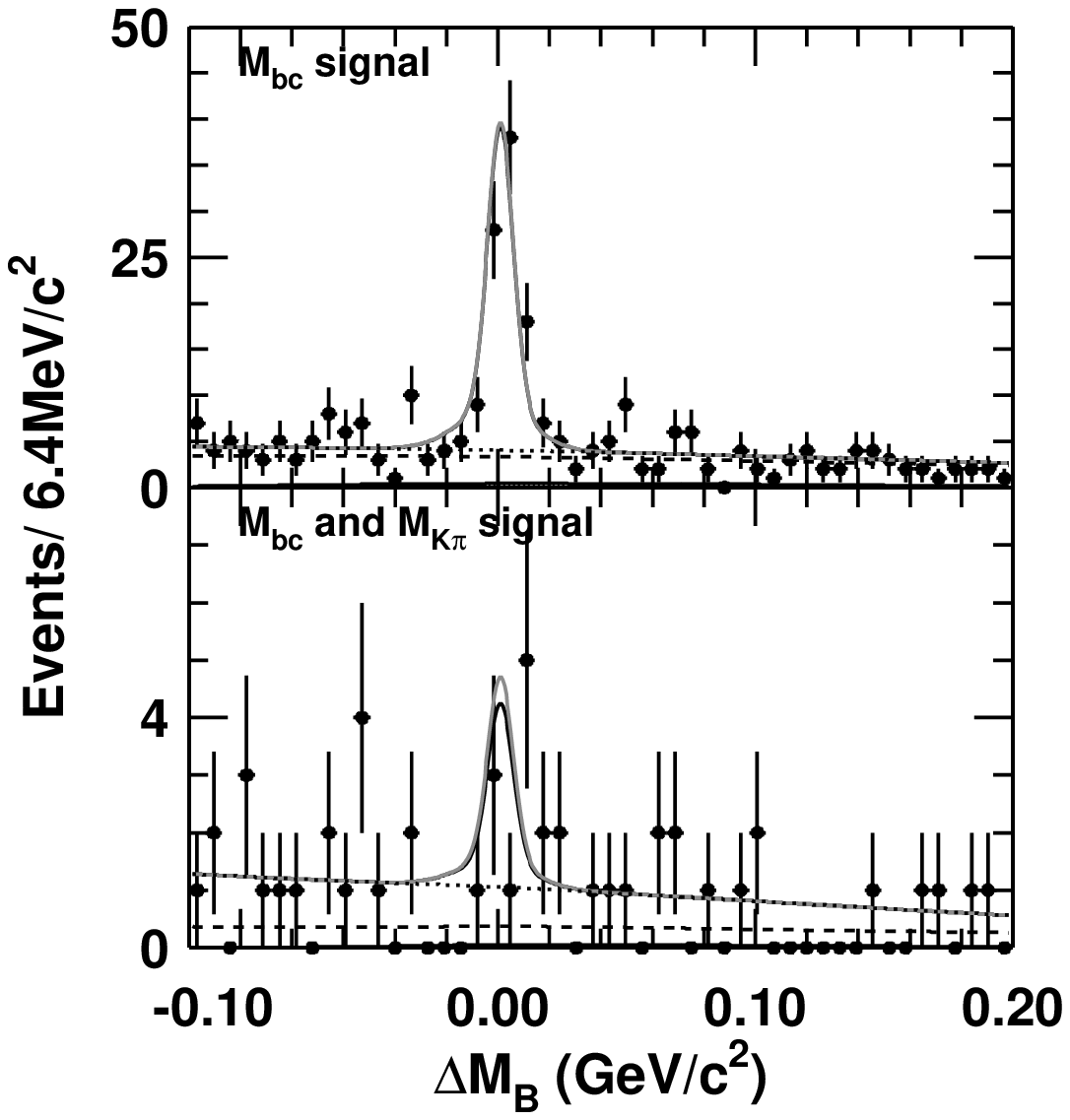}%
\includegraphics[width=0.33\textwidth,height=0.35\textheight]{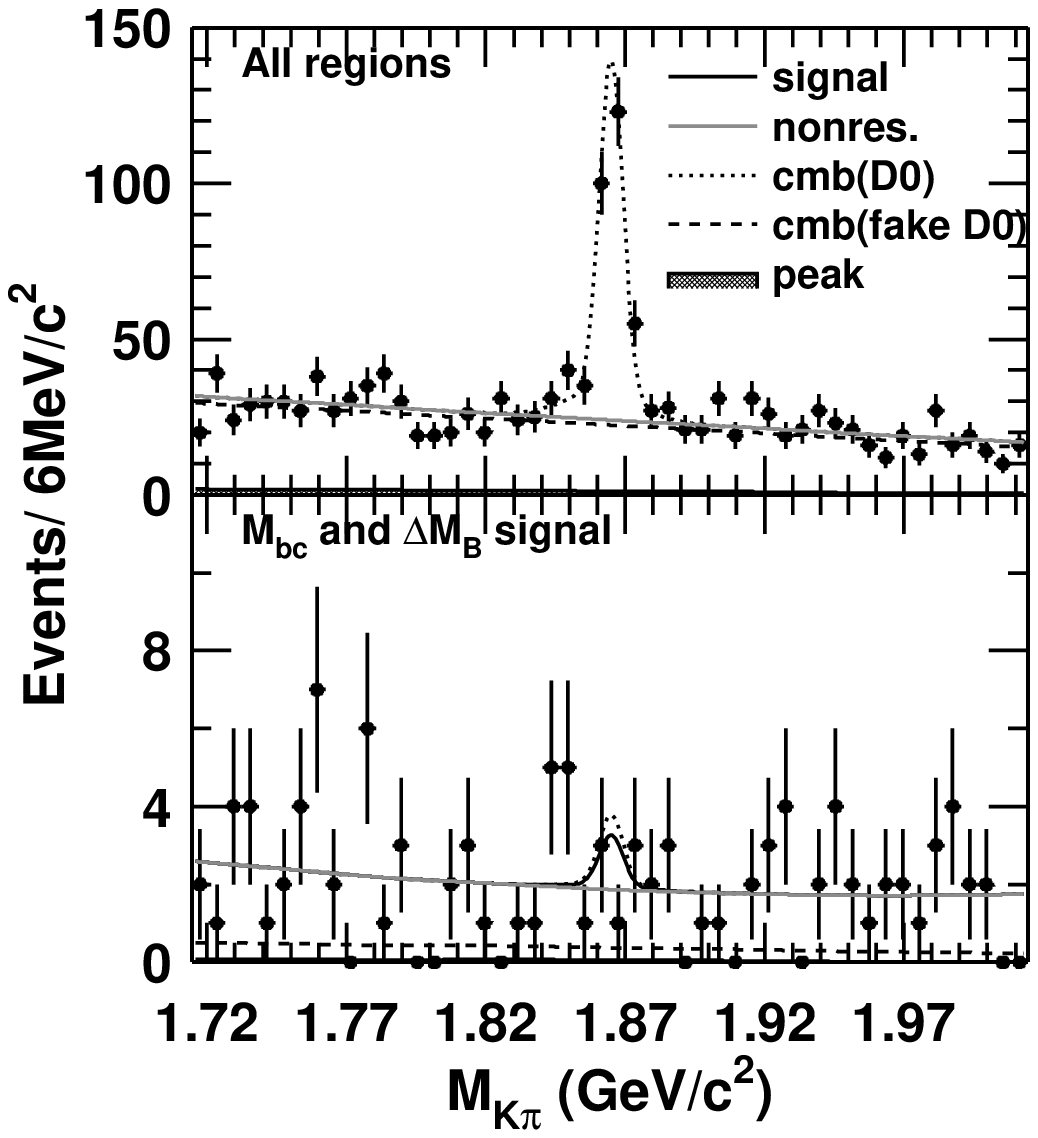}%
\includegraphics[width=0.33\textwidth,height=0.35\textheight]{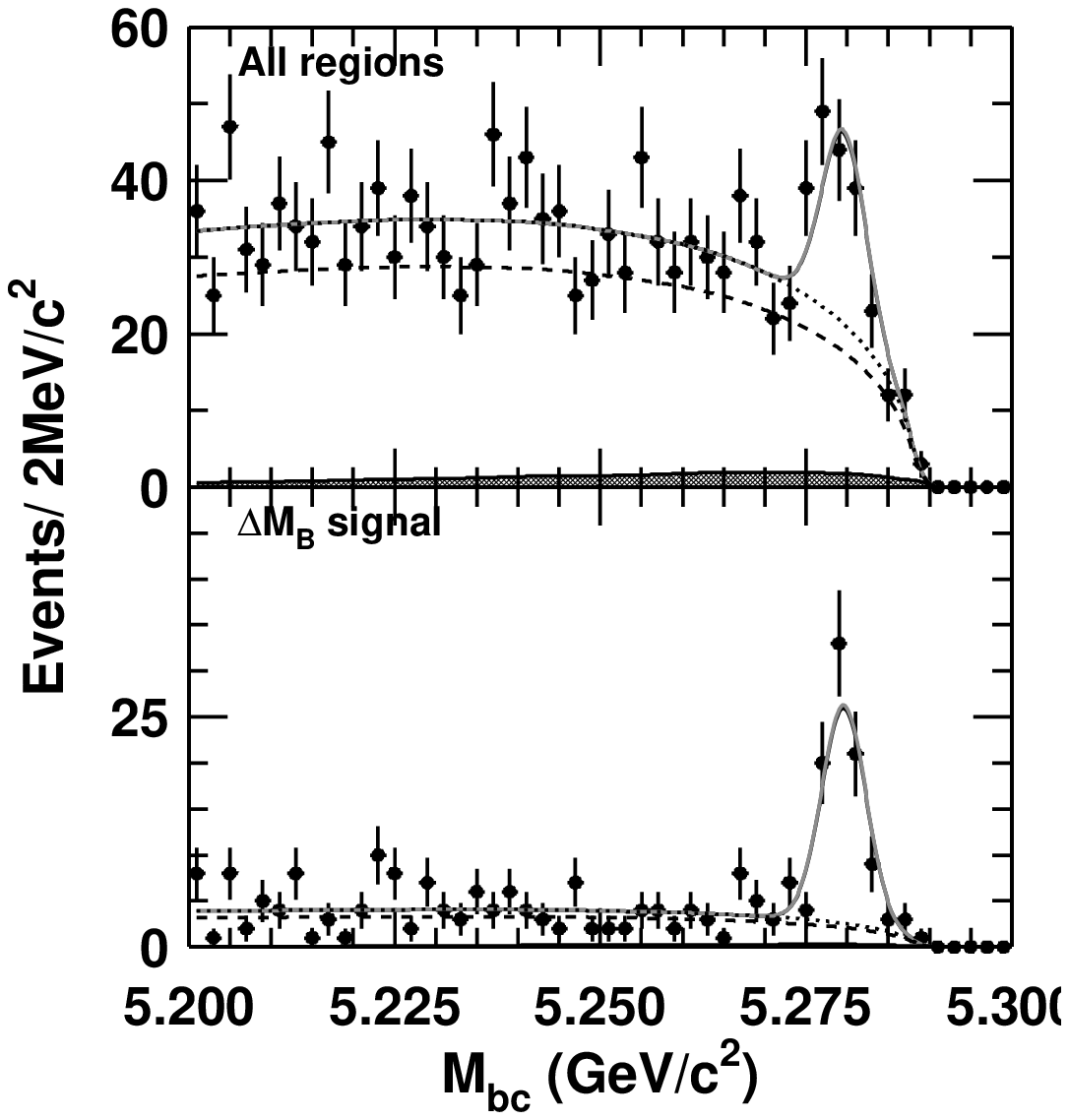}\break
\includegraphics[width=0.33\textwidth,height=0.35\textheight]{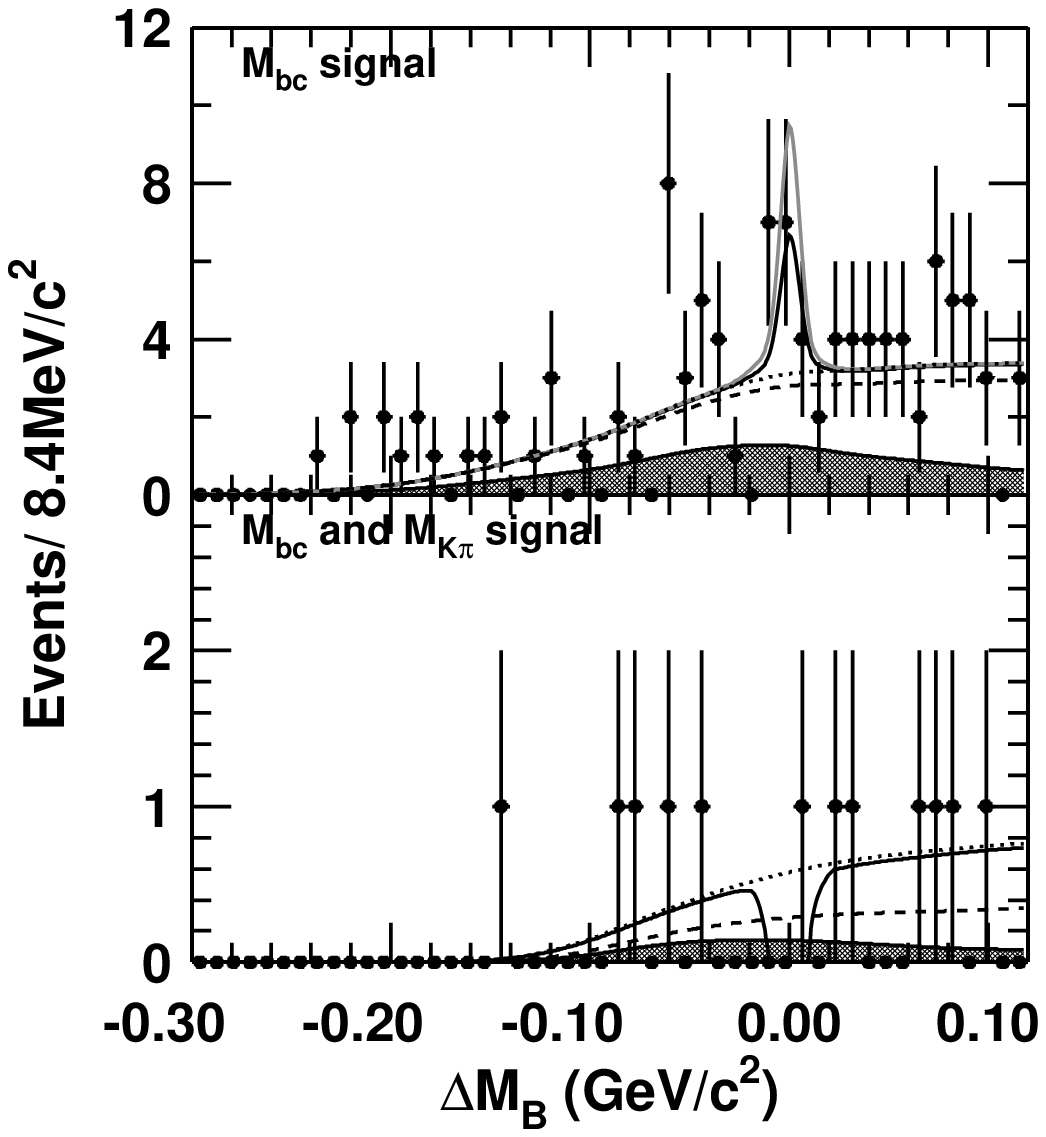}%
\includegraphics[width=0.33\textwidth,height=0.35\textheight]{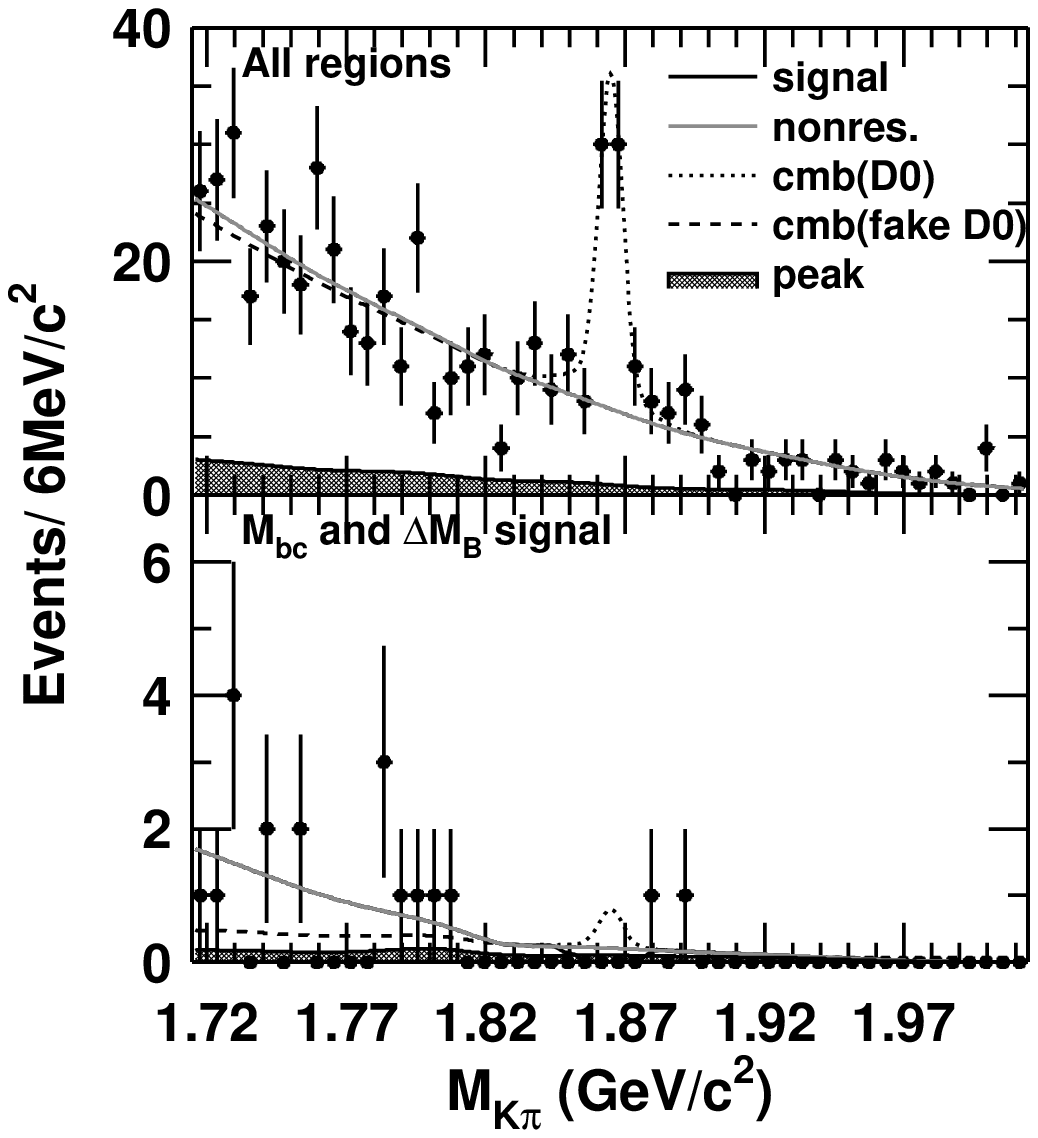}%
\includegraphics[width=0.33\textwidth,height=0.35\textheight]{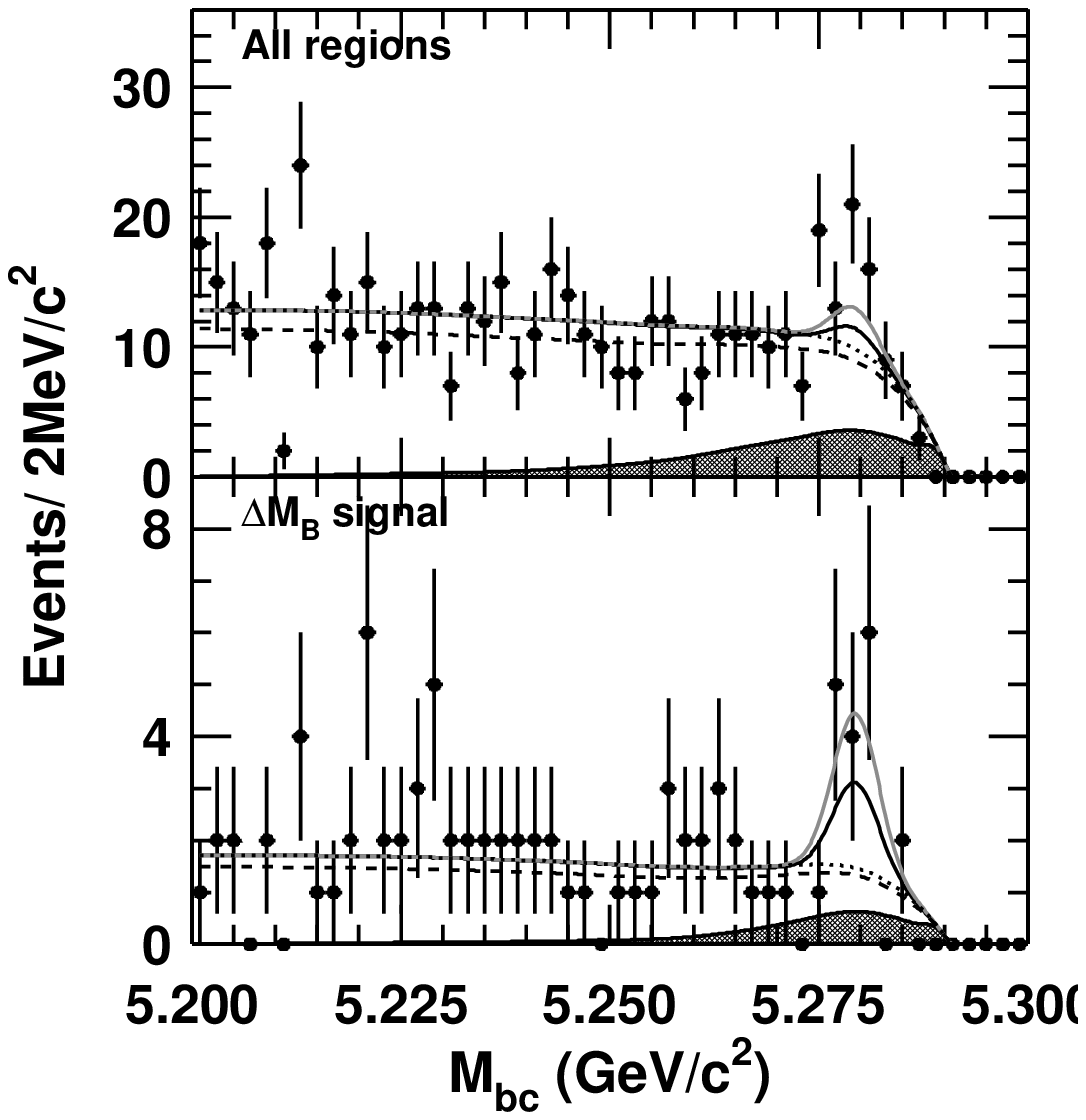}
\caption {\label{proj} $\Dmb$, $\Md$ and $\Mbc$ distributions in all or related variables' signal
regions, as labeled inside each plot, for $\bzdecay$ (top) and $\bpdecay$ (bottom). Superimposed on the
data are projections of the signal and summed background components of the maximum likelihood fit
result.}
\end{center}
\end{figure*}


Figure \ref{sbox} shows ($\Md$, $\Dmb$) scatter plots for candidates in the $\Mbc$ signal region. There
are 11 candidates in the signal region for $\bzdecay$ and one for $\bpdecay$. Table \ref{result}
summarizes the results. The signal yields ($Y$) in the signal box, the expected total background yields
($b$), and their statistical errors are obtained from the maximum likelihood fit. The efficiencies
($\epsilon$) are determined from the signal MC sample with the same event selection
used for the data.
A three-body phase-space model is employed for
$\bpdecay$ decay.   
The $\Dmb$, $\Md$ and $\Mbc$ distributions are plotted in Fig. \ref{proj}.
Also shown are projections of the maximum likelihood fit result, which agree well
with the data. No significant signals are found. We determine 90\% C.L.
upper limits for the signal yield ($Y_{90}$) 
and branching fraction (${\cal B}$) from the observed number of candidates
($n_0$) and the estimated backgrounds ($b$) in the signal box using
the Feldman-Cousins method \cite{FC}. The systematic errors due to the
uncertainties of signal detection efficiency and background yield, elucidated
below, are taken into account \cite{Conrad}.
The decay branching fractions ${\cal B}(\jpsi \to l^+l^-)$ and
${\cal B}(\dz \to K^- \pi^+)$ are taken from the world averages \cite{pdg}.
The fractions of neutral and charged $B$ mesons produced in $\Upsilon(4S)$
decays are assumed to be equal.

\begin{table}[!hbtp]
\begin{center}
\caption{\label{result} Summary of results: $Y$ and  $b$ are the signal and expected total background
yields in the signal box,
$n_{0}$ is the observed number of candidates in the signal box, $\epsilon$ is  the detection
efficiency, $Y_{90}$ and ${\cal B}$ are the 90\% C.L. upper limits for the signal yield and branching
fraction.}
\begin{tabular}{ccccccc}\hline \hline
Mode & $Y$ &$b$& $n_0$& $\epsilon(\%)$ &$Y_{90}$& ${\cal B}(10^{-5})$\\\hline
 $\bz$& $-1.0^{+1.9}_{-1.1}$ &14.6$\pm$1.2$\pm$0.6&
11&
29.9$\pm$2.8& $<4.0$&$<2.0$ \\
$\bp$  &$-4.7^{+1.5}_{-1.0}$ &2.36$\pm$0.36$\pm$0.21& 1&
$14.9^{+2.8}_{-6.0}$ & $<2.5$ &$<2.5$\\
\hline\hline
\end{tabular}
\end{center}
\end{table}

The systematic errors on the background yields are evaluated by varying each fixed PDF parameter by
$\pm1\sigma$ of the measured error, by increasing the order of polynomial for combinatorial background,
and by changing the $\Md$ PDF of nonresonant background to an exponential function. The changes in the
background yields induced by individual variations are added in quadrature. The systematic errors for
the efficiency (Table \ref{eff}) consist of the uncertainties in tracking efficiency of 4.0\% for $\bz$
and 6.1\% for $\bp$ mode, in particle and lepton identification of 2\% per track, in branching
fractions ${\cal B}(\dz \row K^- \pip)$ of 2.4\% and ${\cal B}(\jpsi\row l^+l^-)$ of 1.7\%, and in MC
statistics of 1.2\%. For $\bp \row \jpsi \dbar \pip$ decay, an additional systematic error of
+13.7\%/-38.3\% due to the three-body phase-space model is assigned to the maximum efficiency variation
among the slices of $M(\jpsi,\dbar)$, $M(\jpsi,\pip)$ and $M(\dbar,\pip)$.

\begin{table}[!hbtp]
\begin{center}
\caption{\label{eff} Summary of the contributions to the systematic uncertainty (\%) on the detection
efficiency.} \vskip 0.12 in
\begin{tabular}{lcc} \hline\hline
Source & $\bz$  &   $\bp$\\ \hline
PID efficiency& $8.0$& $10.0$\\
Tracking efficiency& $4.0$& $6.1$\\
MC statistics& $1.2$& $1.2$\\
$\jpsi$ branching fractions&  $1.7$& $1.7$\\
$\dz$ branching fraction & $2.4$ &$2.4$ \\
3-body decay model& -& $+13.7$/$-38.3$\\ \hline
Total & 9.5& 40.2\\
\hline\hline
\end{tabular}
\end{center}
\end{table}

We also obtain the branching fractions of the corresponding nonresonant decay channels in the
$1.71<M_{K^+\pim}<2.01 $ GeV/$c^2$ region from the yields of the nonresonant components in the fit. The
yields in $\Dmb$ and $\Mbc$ signal region are $80.9^{+10.2}_{-9.5}$ for
$\bz \to \jpsi K^+\pi^-$ and $10.1^{+4.0}_{-3.3}$ for $\bp \to \jpsi K^+\pi^- \pi^+$. The efficiencies
are verified to be the same as for the modes with an intermediate $\dz$ resonance. The systematic
errors are estimated by the same procedure. In $\bp$ decay, we subtract the contribution from $B^+
\rightarrow X(3872)K^+$ [$X(3872) \rightarrow J/\psi \pi^+\pi^-$], which is estimated to be $1.20 \pm
0.33$ candidates by MC simulation with the branching fraction taken from Ref. \cite{choi}. Finally, we
obtain ${\cal B}(\bz \row \jpsi K^+ \pim)=(1.51^{+0.19}_{-0.18}\pm0.15)\times10^{-5}$ and ${\cal B}(\bp
\row \jpsi K^+ \pim \pip)=(3.3^{+1.6}_{-1.3}\pm1.6)\times10^{-6}$ in the limited $M_{K^+\pim}$ region
(where the first errors are stat. and the second are syst.).

In summary, we have performed a search for $\bz \row \jpsi \dbar$ and $\bp
\row \jpsi \dbar \pip$ decays. No signal is observed for either decay mode
and upper limits on the branching fraction at 90\%
C.L. are determined to be
\begin{eqnarray*}
 {\cal B}(\bz \row \jpsi \dbar) &<& 2.0 \times 10^{-5}, \\
 {\cal B}(\bp \row \jpsi \dbar \pip) &<& 2.5 \times 10^{-5}.
\end{eqnarray*}
The results are consistent
with the BaBar results \cite{9} and rule out the explanation of the excess in the low momentum
region of the inclusive $\jpsi$ spectrum as intrinsic charm content at the 1\% level in the $B$ meson.

\begin{acknowledgments}
We thank the KEKB group for the excellent operation of the
accelerator, the KEK cryogenics group for the efficient
operation of the solenoid, and the KEK computer group and
the NII for valuable computing and Super-SINET network
support.  We acknowledge support from MEXT and JSPS (Japan);
ARC and DEST (Australia); NSFC (contract No.~10175071,
China); DST (India); the BK21 program of MOEHRD and the CHEP
SRC program of KOSEF (Korea); KBN (contract No.~2P03B 01324,
Poland); MIST (Russia); MHEST (Slovenia);  SNSF (Switzerland); NSC and MOE
(Taiwan); and DOE (USA).
\end{acknowledgments}


\begin{thebibliography}{999}
\bibitem{conjugate} Inclusion of charge conjugate states is implied throughout this paper.
\bibitem{dmb}The benefit of using $\Dmb$ instead of the energy difference is
described in Ref. \cite{zsl}.
\end{thebibliography}
\end{document}